\documentclass[aps,prl,twocolumn,showpacs,floats]{revtex4}
\usepackage{graphicx}%
\usepackage{epsfig}
\usepackage{amsmath}%
\usepackage{amssymb}%

\def\beq{\begin{equation}}
\def\eeq{\end{equation}}
\def\e{\epsilon}
\def\half{{\small {1}\over {2}}}
\def\bighalf{{1 \over  2}}
\def\ksi{\xi}

\def\pr{{\sl Phys. Rev.}\ }
\def\prl{{\sl Phys. Rev. Lett.}\ }
\begin{document}
\bibliographystyle{prsty}
\title{\Large\bf A unifying model for several two-dimensional phase transitions}
\author{Yonatan Dubi$^1$, Yigal Meir$^{1,2}$ and Yshai Avishai$^{1,2}$}
\affiliation{
$^{1}$ Physics Department, Ben-Gurion University, Beer Sheva 84105, Israel\\
$^{2}$ The Ilse Katz Center for Meso- and Nano-scale Science and
Technology, Ben-Gurion University, Beer Sheva 84105, Israel }
\date{\today}

\begin{abstract}
A relatively simple and physically transparent model based on
quantum percolation and dephasing is employed to construct a
global phase diagram which encodes and unifies the critical
physics of the quantum Hall, "two-dimensional metal-insulator",
classical percolation and, to some extent,
superconductor-insulator transitions. Using real space
renormalization group techniques, crossover functions between
critical points are calculated. The critical behavior around each
fixed point is analyzed and some experimentally relevant puzzles
are addressed.
\end{abstract}
\pacs{71.30.+h,73.43.-f,73.43.Nq,74.20.Mn}  \maketitle
Two-dimensional phase transitions have been a focus of
interest for many years, as they may be the paradigms of second
order quantum phase transitions (QPTs). However, in spite of the
abundance of experimental and theoretical information, there are
still unresolved issues concerning their behavior at and near
criticality, most simply exposed by the value of the critical
exponent $\nu$ (describing the divergence of the correlation
length at the transition). Below we discuss three examples which
underscore these problems. First, in the integer quantum Hall (QH)
effect, which is usually described within the single particle
framework, various numerical studies yielded a critical exponent
$\nu \approx 2.35$ \cite{QHnumerics}, in agreement with heuristic
arguments \cite{milnikov}. Some experiments indeed reported values
around $2.4$ \cite{exp2.4}, while other experiments reported
exponents around $1.3$ \cite{exp1.3}, close to the classical
percolation exponent $\nu_p=4/3$. Even more perplexing, some
experiments claim that the width of the transition does not shrink
to zero at zero temperature \cite{noncritical}, in contradiction
with the concept of a QPT. Additionally, the observation of a QH
insulator \cite{QHinsulator} is inconsistent with the QPT scenario
\cite{assa}. Consider secondly the superconductor (SC) - insulator
transition (SIT), for which theoretical studies suggest several
scenarios. Similar to the QH situation, some experiments yield a
value $\nu \approx 1.3$ \cite{SC1.3}, not far from the value
$\nu\simeq1$, predicted by numerical simulations within the random
boson model, but closer to the classical percolation value. Other
experiments, however, yield $\nu\simeq2.8$ \cite{SC2.8}, while
some experiments claim an intermediate metallic phase
\cite{metallic}. As a third example, consider the recently claimed
metal-insulator transition (MIT) \cite{mit}. The critical exponent
is again close to $1.3$ \cite{meir,dassarma} but the occurrence of
such phase transition is in clear contrast with the scaling theory
of localization.

The basic question which naturally arises is then whether it is
possible to unify these two dimensional phase transitions within a
single theory and thereby resolve some of the problems raised
above. A hint toward an affirmative answer is gained by
experimental indications that percolation plays a key role in both
the QH transition \cite{QHperc} and in the SIT \cite{SIperc}.
Moreover, its relevance to the MIT has been argued theoretically
and observed experimentally \cite{meir,dassarma,xi}. The QH and
the SIT have been treated within a percolation-like model in Ref.
\cite{kapitulnik}, consisting of SC or QH droplets connected via
quantum tunneling, similar in spirit to the model presented in
\cite{model}.

In the present work we address these issues using a physically
transparent picture. A similar approach proved to be quite
successful for developing a model encoding the QH transition which
can also describe some aspects of the SIT \cite{model}.
Introducing a new aspect, decoherence, into the model, we are able
to include the QH transition (or the SIT), classical percolation
and the MIT all within the same  phase diagram. Employing
real-space renormalization group (RSRG) techniques, we first
calculate the critical exponents. Most important, we employ this
phase diagram for understanding the possible critical features,
and expose the physical conditions necessary to observe them. The
crossover between different critical points, which may be explored
experimentally is investigated and concrete experimental
predictions are made.

For the sake of self-consistence, the model used to describe the
QH effect \cite{model}, is briefly explained here. In high
magnetic fields, electrons follow equi-potential lines, which, at
low Fermi energy, are trapped in potential valleys. At zero
temperature, transport through the system is due to quantum
tunneling between such trapped trajectories. Within the model,
this picture is mapped on a lattice such that each closed
trajectory in a valley corresponds to a site, and quantum
tunneling between (nearest neighbor) sites corresponds to a
(quantum) link. As the Fermi energy rises and crosses the saddle
point separating two isolated contours, these two trajectories
merge into one. The transmission between these two sites becomes
perfect, and the link connecting them is considered SC. The QH
transition point occurs when such a trajectory spans the entire
system, which, in the model corresponds to percolation
 of the SC links. Consequently, the QH problem maps onto a lattice, with random tunneling amplitudes between its
sites, and a finite fraction of perfect bonds, namely a mixture of quantum and SC links.
 The critical behavior of the QH is encoded by the scaling behavior near the percolation point of this lattice
model.
 Calculation of the transmission through such a system, using two different approaches, yielded a diverging
localization length, with an exponent $\nu \approx 2.4\pm0.1$, close to
the results of other numerical estimates\cite{QHnumerics}.

The new element introduced into the model is dephasing. A fraction
$q$ of the links is attached to current conserving phase
disrupting reservoirs \cite{buttiker}. Physically, it is a
consequence of finite temperature but can result from other
mechanisms. An electron propagating along such a link enters the
reservoir and loses its phase, before it, or another electron, emerges from
 the reservoir (there is no net current into the reservoir).
 Transport along this link is
therefore incoherent.
Transmission in the presence of such phase breaking resistors can
be treated within the Landauer-Buttiker formalism. The model
parameters are then $p$, the fraction of SC links (i.e quantum
links whose transmission is unity), and $q$, the fraction of
classical, or incoherent   links. Evidently, the fraction of
quantum tunneling links is $1-p-q$. The principal objective is
then to construct a phase diagram in the ($p,q$) plane of
parameters and to identify the various critical transitions in
this plane.

Our main tool will be the RSRG scheme on a square lattice. In
order to gain experience and confidence we first perform
calculations on the line $q=0$ which reduces to our previous,
fully coherent model as a special case.  Rescaling the lattice
  by a factor of 2, each $2\times2$ square  is  mapped  onto a $1\times1$
square (Fig~1a).
 Given the transmission amplitude of each link in the $2\times2$ square, we
evaluate the transmission
  through the whole unit, by attaching leads to its left and right sides (see
  below). This will allow us to follow the distribution of the transmission
  amplitudes from one iteration to another, and eventually
  evaluate the distribution for the full lattice.
    For the classical problem this
    calculation reduces to the calculation of transport through a Wheatstone
     bridge (Fig.~1b), where the symmetry of the square lattice has been
  used. For simplicity we use the Wheatstone bridge geometry for the
   quantum problem, and later verify that the addition of dangling bonds
    does not modify the critical behavior.
\begin{figure}[!h]
\centering
\includegraphics[width=8truecm]{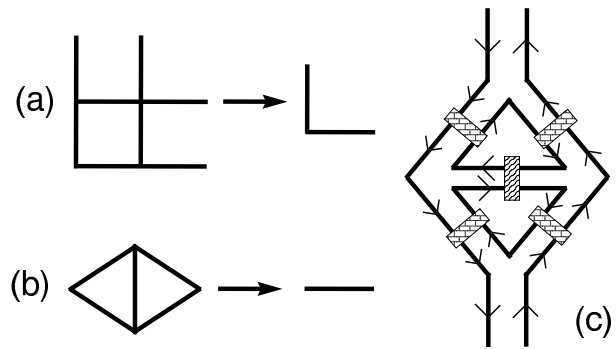}
\caption{\small  The real-space renormalization group process on a
square lattice. (a) The mapping of a $2\times2$ square onto a
$1\times1$ square, resulting
 in reducing the size of the lattice by a factor of $2$. (b) The equivalent
 mapping for which the calculation has been done. For the classical problem
 this equivalence is exact, while for the quantum problem it neglects the
 effects of dangling bonds. Later in the calculation  dangling
  bonds have been included with no observable change in the critical behavior.
(c)  The scattering matrix approach (rotated 90 degrees compared to (b)). At each junction
each edge state
 propagates freely according to its chirality. Scattering is along the links
 (marked as partially filled rectangles across the links). It is easy to see
that
 if there is a percolating path of perfect scatterers (scatterers with unit
transmission),
 then there is a perfect transmission along the whole structure.}
\end{figure}


The actual calculation is carried out using the scattering matrix
approach. Each link carries a left going and a right going
channel. In accordance with the physics at strong magnetic fields
there is no scattering in the junctions (valleys) and the edge
state continues uninterrupted according to its chirality (see
Fig.~1c), while the scattering occurs on the link (saddle point)
itself.
 Each scatterer is characterized by its scattering matrix $S_i$, namely a
transmission
 probability $T_i$ and phases. For each realization of
   transmission probabilities on the larger lattice, we can evaluate exactly the
   transmission through the whole lattice $\hat{T}(\{T_i\},\phi)$ , and thus
follow its
    distribution from one iteration to another, allowing us to determine the
    fixed points and the RSRG flow. If one starts with a distribution of
scattering
    matrices whose phases are uniformly distributed, they remain so, and
    thus we will be only interested in the scaling of the distribution of the
    transmission probabilities,
\begin{eqnarray}
G^{\{n\}}(T) = \int \ldots \int \   \delta
\left(T-\hat{T}(T_1,\ldots,T_5,\boldsymbol{\phi})\right) &\times& \nonumber \\
\times\ G^{\{n-1\}}(T_{1}) \cdots G^{\{n-1\}}(T_{5})\  dT_{1} \cdots dT_{5}\
d\boldsymbol{\phi},& &
\label{recur}
\end{eqnarray}
where $\boldsymbol{\phi}$ stands for all the independent phases.


 The initial distribution can be written as \beq
G^{\{0\}}(T) = p\ \delta(T-1) + (1-p) \hat{G}^{\{0\}}(T) ,
\label{g0} \eeq where $p$ is determined by the fraction of
saddle-point energies that are below the Fermi energy,
$p(\e_F)=\int_{-\infty}^{\e_F}F(\e) d\e$, with $F(\e)$ the
distribution of
 the saddle point energies.
$\hat{G}^{\{0\}}(T)$ is determined by the relation between the
transmission and the saddle-point energy and by $F(\e)$. The
initial distribution $\hat{G}^{\{0\}}(T)$ does not affect the
critical behavior, as the distribution always flows toward one of
three possible fixed-point distributions.

Since the probability that the transmission is unity is determined by the
classical
percolation probability, one finds
\beq
G^{\{n\}}(T) = P_n(p) \delta(T-1) + \left(1-P_n(p)\right) \hat{G}^{\{n\}}(T,p) ,
\eeq where $P_n(p)$ is the classical percolation probability after
$n$ iterations, and $\hat{G}^{\{n\}}$ can be exactly expressed in
terms of $\hat{G}^{\{n-1\}}$. Clearly, if $p<p_c$, $P(p)$ flows
toward zero, while for $p>p_c$ it flows to unity. At $p=p_c$ the
distribution $\hat{G}$ flows toward a two-peak fixed distribution,
 similar to previously obtained results for the
QH transition \cite {distribution}.
 The critical behavior can be deduced by investigating the length
dependence of the averaged transmission near the critical point
and its collapse using
$T(p,L)=T[L/\ksi(p)]$, where $\ksi(p)$ is an energy (or
concentration) dependent localization length. We find that
$\ksi(p)$ diverges at $p_c=1/2$ with an exponent $\nu=2.4\pm0.1$,
consistent with previous numerical calculations for the QH
transition \cite{QHnumerics}.



Having demonstrated the power of the RSRG technique for $q=0$ (no
dephasing) we now treat the model at an arbitrary point ($p,q$).
Qualitatively, several phase transitions can be identified. For
$q=0$ the QH transition is recovered as demonstrated above. For
$p+q=1$ (no quantum links) the point $q=p_c=\half$ describes the
classical conductor-superconductor percolation transition. For the
case of $p=0$, all the links are either quantum (with $T<1$) or
classical. This is the model suggested in Refs. \cite{meir,xi} to
describe the "apparent" metal-insulator transition in two
dimensions.  The flow lines can be determined without the full
quantum calculations by noticing that the rescaled cell will be SC
if a cluster of SC links percolate, while it will be an incoherent
metal if  there is percolation of classical links, and no
percolation of SC ones. These conditions define the RSRG equations
for the quantities $p$ and $q$,
\begin{eqnarray}
 p ' & =&  2\,{\left( 1 - p \right) }^3\,p^2 + 8\,{\left( 1 - p \right) }^2\,
p^3 +
  5\,\left( 1 - p \right) \,p^4 + p^5 \\
 q ' &=&  q\left\{10p^4-20p^3\left(1-q\right)  +
    2p\left(1-q\right) \left[2+5\left(1-q \right) q\right]  \right. \nonumber \\
&+&   \left.  q\left( 2 + 2q - 5q^2 + 2q^3 \right)  + 2p\left( 2 + 3q - 10q^2 +
5q^3 \right)  \right\} \nonumber
\label{rg}
\end{eqnarray}

\begin{figure}[!h]
\centering
\includegraphics[width=7truecm]{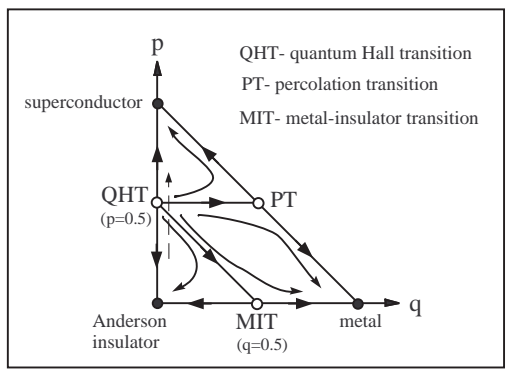}
\caption{\small The phase diagram of the model in term of the
parameters $q$, the concentration of a perfect-transmission links
and $p$, the probability of incoherent links. The phase diagram
displays three different phase transitions between the three
different phases. The broken  line represent a possible
experimental trajectory leading to a finite metallic phase.}
\end{figure}


The full phase diagram thus contains three limiting phases, an
Anderson insulator, a superconductor (or a QH phase) and an
incoherent metal, and is depicted in Fig.~2. The fact that for
$p>\half$ one gets a QH phase is straightforward, as once there is
percolation of SC links there is an edge state propagating without
backscattering through the system. The transition between an
Anderson insulator and an incoherent metal is more intriguing.
Once $p+q>\half$ an electron can propagate freely from one side of
the system to the other, but necessarily goes through incoherent
scattering, and thus this phase is an incoherent metal. For
$p+q<\half$, however, there is no trajectory that allows the
electron to traverse the system without quantum tunneling. These
tunneling events, however, are intertwined with incoherent
scattering, and thus
 the conductance of the system will be of the order of $\exp(-L_{\phi}/\ksi)$,
where
 $L_{\phi}\sim q^{-1/2}$ is the average distance between incoherent scattering events.
As $q$ decreases,
 the conductance of the system goes to zero, ending up as an Anderson
insulator. For finite
 $p$, however, the existence of incoherent scattering is not expected to affect
the quantization
 of the Hall conductance \cite{efrat}. Thus one may characterize the phase with
finite $p$ and
 $q$, but with $p+q<\half$ as a "Hall insulator". In agreement with previous
treatments of
  this phase, we also conclude that it only exists for finite dephasing
($q>0$), as for $q=0$
  the system flows to an Anderson insulator. Further investigation
  of this phase will be reported elsewhere.

Beyond the characterization of the different phases, the model
allows a direct calculation of the crossover between them.
 The line at $p=\half$ allows, as a function of $q$, to
study the crossover between the quantum SIT (or the QH transition)
to the classical superconductor-conductor transition. Following
the RSRG flow along this line, we find that the average
transmission obeys
\begin{equation}
T(q,p=\bighalf) = T_0 \left( 1 + \beta
\left({L\over{L_{\phi}(q)}}\right)^{0.91} \right) , \label{co1}
\end{equation}
With $L_{\phi}(q)$ defined above and $\beta$ some nonuniversal constant. This
function
correctly reproduces the two fixed point behaviors, $T = T_0$ at the QH
transition and
$T\sim L^{t/\nu_p}$ at the classical percolation critical point. The values of
 $t$, the resistance critical exponent, and $\nu_p$ obtained separately by RSRG are
 $1.33$ and $1.42$, respectively, giving $t/\nu_p \sim 0.93$, in good agreement
with
 the limiting behavior of the crossover function (\ref{co1}). The effect of
incoherent
 scattering on the critical behavior may explain the different critical
exponents observed
 in QH and in SITs, which mostly agree with one of the critical exponents
associated with
 these two critical points.

The QH to the MIT crossover has been demonstrated  experimentally
\cite{hanein}. The present formalism allows us to study this
crossover theoretically, by following the conductance along  the
line $p+q=\half$. The dependence of the transmission on length,
for different values of $q$ along this line is depicted in Fig.~3.
There is a clear crossover from a constant (QH behavior) to $T\sim
L^{t/\nu_p}$ (classical behavior), as before. The length scale
determining this crossover (see inset of Fig.~3) is found to be
the percolation correlation length, namely for $L<\ksi_p$ the
system behaves quantum mechanically, and classically for
$L>\ksi_p$.

\begin{figure}[!h]
\centering
\includegraphics[width=8.0truecm]{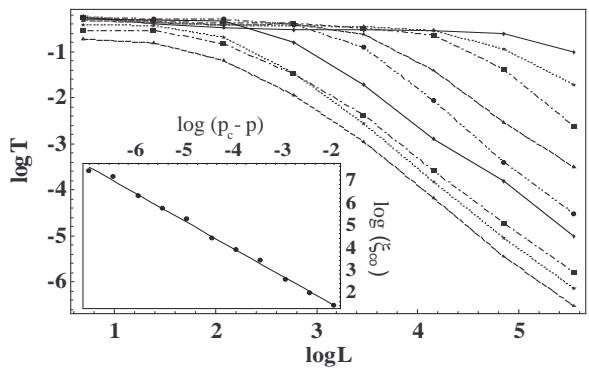}
\caption{\small Crossover from the QH critical
point to the MIT: transmission as a function of length (on a
log-log scale) for different values of $(p_c-p)$ , demonstrating
the crossover from the QH behavior (length-independent
transmission) to a classical power-law dependence. Inset: The
crossover length $\ksi_{co}$ as a function of the $(p_c-p)$ on a log-log
scale, with a slope $\nu=1.23$, close to the percolation
correlation exponent. }
\end{figure}


 All these observations have clear experimental
relevance. As mentioned above, the QH to the MIT
 crossover has been investigated experimentally, but with no detailed
investigation of the
 critical behavior. We thus predict that the critical exponent along this line
will be the
 classical percolation exponent, $\nu_p$, and not the QH exponent, with two relevant length scales
 for any finite system. Similarly,
one can experimentally
 investigate the quantum SIT to the classical superconductor - conductor
transition, for different
 temperatures. The theory predicts that once the dephasing
length becomes
 smaller than the system size (i.e. $q$ becomes nonzero), the critical behavior
will correspond
 to classical percolation. Moreover, we predict that the conductance at the critical
point will
  vary from its universal value at the quantum transition to a length
dependence value
  according to Eq.(\ref{co1}).

  Another result of the proposed phase diagram is the existence of an
intermediate metallic regime
  in the QH and in the SITs. Imagine changing a physical parameter (e.g.
density or magnetic
  field) along the broken line in Fig.~2. Then there is one transition from an
insulator
   to a metal, and then from a metal to a QH liquid (or a superconductor),
consistent with
   experimental observations for the QH transition\cite{noncritical} and the
SIT \cite{metallic}.
  The critical behavior of both these transitions is determined by classical
percolation. Thus the
  theory predicts that the QH critical behavior can only be observed in system
where these
  transitions coalesce into a single transition. This can be tuned, for
example, by lowering
  the temperature and changing the dephasing length. Thus the theory can
explain the different
  exponents observed experimentally and predicts a crossover between the
different critical
  behaviors as a function of temperature.

  It should be noted that some aspects of this phase diagram are not dissimilar
to the
  one presented in Ref.\cite{dissipation}. Here, however, we do not invoke any
zero temperature
  dissipation. Rather, if, as expected, the dephasing length diverges at
$T\rightarrow0$
   (i.e. $q\rightarrow0$) the theory predicts only two stable phases - the QH
(or superconducting)
   phase and the Anderson insulator phase.

We would like to acknowledge fruitful discussions with A. Aharony
and O. Entin-Wohlman. This research has been funded by the ISF.

\end{document}